\documentclass[twocolumn]{revtex4-2}
\usepackage[utf8]{inputenc}
\usepackage{graphicx}
\usepackage{stackengine}
\usepackage{xcolor}
\usepackage[colorlinks,allcolors=cyan!70!black, breaklinks=true]{hyperref}

\usepackage{tcolorbox}
\usepackage{algpseudocode}
\usepackage{mathtools}

\usepackage[english]{babel}
\renewcommand{\selectlanguage}[1]{}

\usepackage[inkscapearea=page]{svg}
\usepackage{lipsum}

\DeclarePairedDelimiter\autobracket{(}{)}
\newcommand{\PA}[1]{\autobracket*{#1}}
\begin{document}

\title{Apparent phase transitions and critical-like behavior in
  multi-component mixtures}

% Are phases appropriate for cells? -- Fluctuation dominated regimes
% in finite-size systems}

\author{Felix Herrmann}

\affiliation{Max Planck Institute for Polymer Research, Ackermannweg
  10, Mainz 55128, Germany}

\author{Burkhard Dünweg}

\affiliation{Max Planck Institute for Polymer Research, Ackermannweg
  10, Mainz 55128, Germany}

\author{Martin Girard}

\email{martin.girard@mpip-mainz.mpg.de}

\affiliation{Max Planck Institute for Polymer Research, Ackermannweg
  10, Mainz 55128, Germany}

\begin{abstract}
  Liquid-liquid phase separation has recently emerged as an important
  topic in the context of cellular organization. Within this context,
  there are multiple poorly understood features; for instance hints of
  critical behavior in the plasma membrane, and how homeostasis
  maintains phase separation. In this paper, using statistical
  mechanics, we show that finite size effects in multicomponent
  mixtures can induce the system to behave as-if it were near a
  critical point, which we term apparent transitions. The apparent
  transition temperature is naturally driven towards the ambient
  temperature of the system.
\end{abstract}
\maketitle

\section*{Main}
Liquid--liquid phase separation (LLPS) can nowadays be considered as
fairly well understood --- as far as those aspects are concerned that
can be treated within the framework of equilibrium statistical physics
of infinite systems that are composed of a small number of molecular
species. In particular, polymer physics has highlighted the unmixing
propensity of high-molecular weight systems, as a direct consequence
of the loss of translational entropy due to molecular connectivity. In
recent years, LLPS has nevertheless found renewed strong interest
because there are strong indications that it is of major importance
for the functioning of biological
systems~\cite{alberti_liquidliquid_2019,
  schaffert_post-translational_2020, burns_miscibility_2017,
  shaw_critical_2021, alberti_considerations_2019, shin_liquid_2018,
  banani_biomolecular_2017, lafontaine_nucleolus_2021}. Here the
physics is, to a large extent, still poorly understood, since the
situation is in many decisive aspects quite different from the
``classical'' scenario: Firstly, the systems are typically driven and
not in thermal equilibrium; secondly, the number of different
molecular species is huge; and thirdly the systems are fairly small.

The present paper is a first attempt to make some progress in better
understanding of LLPS in biological systems. As a complementary
approach to existing studies~\cite{sear_instabilities_2003,
  graf_thermodynamic_2022, thewes_composition_2023,
  shrinivas_phase_2021, jacobs_predicting_2013, jacobs_phase_2017}, we
here focus only on two aspects, namely the finite size and the large
number of molecular species. Non-equilibrium physics is deliberately
ignored, in order to keep the analysis simple and also in order to
assess the importance of the other two ingredients. We will therefore
focus on a very simple statistical--mechanical model, which to a
significant extent can be treated analytically, and try to understand
its properties well. The simplifications that we make are so severe
that a mapping onto real biological systems should probably not be
attempted; however, we believe our model and findings are interesting
in their own right, just from the point of view of statistical
physics, and may perhaps help us in improving our understanding of
biological systems at least somewhat.

\subsection*{Model: Definition and Basic Considerations}

We consider a $d$--dimensional simple cubic lattice of $L \times L
\times L \dots \times L = L^d$ sites with periodic boundary
conditions, where often we choose $d = 2$, inspired by LLPS observed
in membranes. We enumerate the lattice sites with an index $i$, $i =
1, \ldots, L^d$, and we assume that every site is occupied by a
molecule (no voids). Interactions are confined to nearest neighbors
only, and $\langle i j \rangle$ denotes a nearest--neighbor pair
composed of the sites $i$ and $j$.

We now assume that for each molecule we can identify a vector of
quantitative (and suitably non-dimensionalized) properties, such as
size and shape parameters, etc., $\vec{s}$ (``descriptor''), such that
this information is sufficient to uniquely find the interaction energy
$f \left(\vec{s}, \vec{s}^\prime \right)$ of a pair of molecules
forming a nearest--neighbor bond. We now make two strongly simplifying
assumptions: (i) We assume that the interactions are translationally
invariant in descriptor space, such that
$f \left(\vec{s}, \vec{s}^\prime \right) = f \left( \left\vert \vec{s}
    - \vec{s}^\prime \right\vert \right)$ --- or that such a property
holds, to sufficient accuracy, at least approximately, after suitable
re--parametrization. (ii) We assume, for simplicity, that the
descriptor space is one--dimensional, such that the interaction is
given by
$f \left( s, s^\prime \right) = f \left( \left\vert s - s^\prime
  \right\vert \right)$. Although this latter assumption may be lifted
without major technical difficulties (see SI), we keep it in the
development for ease of notation, and for constructing a unique and
well--defined model. For simplicity, we assume that we have a
continuous spectrum of descriptor values $s \in (-\infty, +\infty)$
available, which may be viewed as looking at a system which is able to
host an infinite number of molecular species. In practice, of course,
the system will host (at most) ``only'' $L^d$ different descriptor
values.

If we now assume that we have specified a function $f$, the
internal energy of a system of molecules arranged on the lattice
is then simply given by
\begin{equation}
U = \sum_{\langle i j \rangle}
f \left( \left\vert s_i - s_j \right\vert \right),
\label{eq:SimpleInteractionEnergy}
\end{equation}
where $s_i$ is the descriptor of the molecule located on
site $i$. It should be noted that this Hamiltonian
exhibits no quenched disorder whatsoever, which is
of course a substantial simplification.

Based upon Eq.~\ref{eq:SimpleInteractionEnergy} one could, in
principle, now directly run a Monte Carlo (MC) simulation in the
canonical ensemble, where, starting from some initial configuration,
one randomly picks a pair of molecules (or descriptors on the
lattice), attempts to exchange them, and accepts or rejects that swap
via a standard Metropolis criterion based upon the desired Boltzmann
weight $\exp \left( - \beta U \right)$, where $\beta = 1 / (k_B T)$,
$k_B$ denoting the Boltzmann constant and $T$ the absolute
temperature. This procedure is repeated again and again, until the
system has relaxed into equilibrium, from when on the sampling of
properties begins.

This, however, runs into both a technical and a conceptual
difficulty. The technical difficulty is that the exchange algorithm is
quite slow and difficult to parallelize, such that substantial
computer resources would be needed.
% We therefore do not present any such simulations in the present paper.
The conceptual difficulty is that the initial configuration must be a
random sample of some descriptor values assigned to the lattice
sites. The set of values then remains conserved during the simulation,
while the assignment to the lattice sites evolves. This immediately
rises the question of the underlying probability distribution of the
random initial configuration, and also points to the need to average
over initial conditions in order to obtain meaningful results.

Statistical physics as such is unable to tell us anything about the
probability density of $s$, since this is a result of the frequency of
occurence of the various molecular species, which in turn is a result
of biology. We can however give it a name --- we call it the
``molecular probability density'' $\Omega (s)$, with
$\int ds \, \Omega (s) = 1$ --- and make model assumptions about it
(see below).

In order to formulate the underlying statistical physics, we need some
notation, which may look somewhat tedious and obvious, but is in our
opinion needed for clarity. A configuration of the system is specified
by the values $s_1, s_2, \ldots, s_{L^d}$, which we can combine to
form a large vector
$\vec{S} = \left( s_1, s_2, \ldots, s_{L^d} \right)$. Within the
canonical ensemble (i.~e. within the framework of a
descriptor--exchange MC simulation) we know that any permutation of
vector components will result in a new configuration that is also part
of the state space that is accessible to the system. In other words: A
permutation matrix $\Pi$ applied on the initial vector $\vec{S}_0$
will result in a new element of the state space
$\vec{S} = \Pi \vec{S}_0$; considering all possible permutations will
exhaust the state space completely. An observable $A$ is nothing but
a function defined for any vector $\vec{S}$,
$A = A \left( \vec{S} \right)$, and its thermal equilibrium average
within the framework of a descriptor--exchange simulation is given by
\begin{eqnarray}
  &&
  \left< A \right>_c
  \left( \vec{S}_0 \right)
  \\
  \nonumber
  & = &
  \frac{
  \sum_{\Pi}
  A \left( \Pi \vec{S}_0 \right)
  \exp \left[ - \beta U \left( \Pi \vec{S}_0 \right) \right]
  }{
  \sum_{\Pi}
  \exp \left[ - \beta U \left( \Pi \vec{S}_0 \right) \right]
  } ,
\end{eqnarray}
where the notation tries to emphasize that the value depends on the
initial configuration $\vec{S}_0$, or, more precisely, on the
associated set composed of $\vec{S}_0$ and all its permutation
images. The subscript $c$ indicates an average in the canonical
ensemble.  It is clear that this may be viewed as the definition of a
new ``observable'' $\left< A \right>_c \left( \vec{S} \right)$, which
is permutation invariant by construction.

In order to remove the dependence on the arbitrarily chosen initial
condition (or associated set), one somehow needs to average over the
full state space.  To this end, we assume that each molecular species
(or each descriptor value) has its own independent particle reservoir,
such that the molecular probability density to find a configuration
$\vec{S}$ is given by
$\tilde{\Omega} \left( \vec{S} \right) = \Omega \left( s_1 \right)
\times \Omega \left( s_2 \right) \times \ldots \times \Omega \left(
  s_{L^d} \right)$. With
$d \vec{S} = ds_1 \, ds_2 \, \ldots \, ds_{L^d}$ we thus have
$\int d \vec{S} \tilde{\Omega} \left( \vec{S} \right) = 1$. As a side
remark, note that for each permutation $\Pi$ we have
$\tilde{\Omega} \left( \Pi \vec{S} \right) = \tilde{\Omega} \left(
  \vec{S} \right)$.  Conceptually, the most straightforward ``full''
average for $A$ would therefore be
\begin{equation}
  \left< A \right>_q =
  \int d \vec{S} \, \tilde{\Omega} \left( \vec{S} \right)
  \left< A \right>_c \left( \vec{S} \right) ,
\end{equation}
where the subscript $q$ indicates that this type of averaging
is similar to the so--called ``quenched averages'' known from
the theory of disordered systems. In a simulation, this means
that one needs to generate a large number of independent
trajectories, each with a different initial condition generated
according to the probability density $\tilde{\Omega}$, which are
then averaged over.

In contrast to this type of average, we may also define an
``annealed'' average $\left< A \right>_a$, where the probability
density $\tilde{\Omega}$ is directly combined with the Boltzmann
weight:
\begin{eqnarray}
  &&
  \left< A \right>_a
  \\
  \nonumber
  & = &
  \frac{
  \int d \vec{S} \, \tilde{\Omega} \left( \vec{S} \right)
  \sum_{\Pi}
  A \left( \Pi \vec{S} \right)
  \exp \left[ - \beta U \left( \Pi \vec{S} \right) \right]
  }{
  \int d \vec{S} \, \tilde{\Omega} \left( \vec{S} \right)
  \sum_{\Pi}
  \exp \left[ - \beta U \left( \Pi \vec{S} \right) \right]
  }
  \\
  \nonumber
  & = &
  \frac{
  \sum_{\Pi}
  \int d \vec{S}  \, \tilde{\Omega} \left( \vec{S} \right)
  A \left( \Pi \vec{S} \right)
  \exp \left[ - \beta U \left( \Pi \vec{S} \right) \right]
  }{
  \sum_{\Pi}
  \int d \vec{S} \, \tilde{\Omega} \left( \vec{S} \right)
  \exp \left[ - \beta U \left( \Pi \vec{S} \right) \right]
  }
  \\
  \nonumber
  & = &
  \frac{
  \sum_{\Pi}
  \int d \vec{S} \, \tilde{\Omega} \left( \vec{S} \right)
  A \left( \vec{S} \right)
  \exp \left[ - \beta U \left( \vec{S} \right) \right]
  }{
  \sum_{\Pi}
  \int d \vec{S} \, \tilde{\Omega} \left( \vec{S} \right)
  \exp \left[ - \beta U \left( \vec{S} \right) \right]
  }
  \\
  \nonumber
  & = &
  \frac{
  \int d \vec{S} \, \tilde{\Omega} \left( \vec{S} \right)
  A \left( \vec{S} \right)
  \exp \left[ - \beta U \left( \vec{S} \right) \right]
  }{
  \int d \vec{S} \, \tilde{\Omega} \left( \vec{S} \right)
  \exp \left[ - \beta U \left( \vec{S} \right) \right]
  } ,
\end{eqnarray}
where we have exploited the permutation invariance of $d \vec{S}$
and of $\tilde{\Omega}$.

We now may define a chemical potential $\mu (s)$ for the
descriptor value $s$ via
\begin{equation}
  \mu (s) = k_B T \ln \Omega (s),
\end{equation}
which allows us to write
\begin{equation}
  \tilde{\Omega} \left( \vec{S} \right)
  \exp \left[ - \beta U \left( \vec{S} \right) \right]
  =
  \exp \left[ - \beta U_{eff} \left( \vec{S} \right) \right]
\end{equation}
with the Hamiltonian
\begin{equation}
  U_{eff} \left( \vec{S} \right) =
  U \left( \vec{S} \right) - \sum_i \mu (s_i) ,
\end{equation}
such that
\begin{eqnarray}
  &&
  \left< A \right>_a
  \\
  \nonumber
  & = &
  \frac{\int d \vec{S}
  A \left( \vec{S} \right)
  \exp \left[ - \beta U_{eff} \left( \vec{S} \right) \right]
  }{
  \int d \vec{S}
  \exp \left[ - \beta U_{eff} \left( \vec{S} \right) \right]
  } .
\end{eqnarray}

We have thus defined three different procedures to define the average
of $A$, and one may view this as the definition of three different
statistical--mechanical ensembles: The ``canonical'' ensemble,
the ``quenched'' ensemble, and the ``annealed'' (or grand--canonical)
ensemble.

Usually, statistical physics is interested in the thermodynamic limit
$L \to \infty$, and in the leading--order corrections in terms of the
ratio $\xi / L$, where $\xi$ is the descriptor--descriptor correlation
length in an infinite system. In other words, one typically assumes
$\xi \ll L$, which may, at best, be violated in a fairly narrow region
in the vicinity of a critical point. In contrast, we are here
interested in the opposite limit, where $\xi \gg L$, which may be
achieved not only by approaching a critical point, but also (as we
will see below) by simply choosing a sufficiently low temperature. The
most important consequence of this is the \emph{non--equivalence of
  ensembles}: While for $\xi \ll L$ one has
$\left< A \right>_c \simeq \left< A \right>_q \simeq \left< A
\right>_a$, and the choice of ensemble is mainly a matter of practical
convenience, this is no longer the case here --- rather, one must
expect that the system will behave quite differently, depending on
what ensemble is assumed, or, in more physical terms, what
\emph{constraints} the system is subjected to. In particular, we will
see that a system which, in the grand--canonical ensemble for
$L \to \infty$, does not show any phase transition whatsoever,
\emph{will}, in the canonical ensemble, exhibit features that
\emph{resemble} a phase transition --- although it is of course clear
that the finite system size precludes the presence of a true
thermodynamic singularity.

Throughout this paper, we will investigate statistical
distributions. The natural quantities to characterize these are
central moments. For a univariate random variable $x$ with probability
density $P(x)$, and some arbitrary function $\phi(x)$, the expectation
value of $\phi$ is given by
\begin{equation}
  \left< \phi \right>_P = \int_{-\infty}^{\infty} dx \, P(x) \phi(x),
\end{equation}
and this allows us to define the corresponding $n$-th centered moment
as
\begin{equation}
m_n(P) = \langle \left( x - \langle x \rangle_P \right)^m \rangle_P .
\label{eq:moment}
\end{equation}
The second centered moment coincides with the variance. Following
standard statistical definition, the skew is given by
$\gamma(P) = m_3(P) / m_2(P)^{3/2}$, and the kurtosis by
$\kappa(P) = m_4(P) / m_2(P)^2$.

We consider two different kind of averages in \eqref{eq:moment}. When
used without any label, i.e. $\langle\cdot\rangle$, the average refers
to ensemble averages, as previously defined. An ensemble average
itself is not a random variable, and consequently, the same is true
for any centered moment derived from it.

By opposition to ensemble average, and as in
\cite{girard_finite-size_2021}, we also use volume-averages, which we
denote by $\langle\cdot\rangle_L$. In general, the volume average is a
random variable, to which we can assign a distribution and centered
moments. The volume average coincides with the canonical
average. Consequently, in the canonical ensemble, all moments of the
volume-average are zero.

In the context of this article, the probability density of $\vec{S}$,
which we label $\mathcal{D}$, plays a central role. For clarity, we
give special notation to the moments of
$\mathcal{D}(\vec{S})$. Namely, the volume-average variance is labeled
$\sigma_L^2$ and the volume-averaged skew is labeled $\gamma_L$
(precise definitions see below).

\subsection*{Gaussian system}

We first consider a system in the grand canonical ensemble, where we
assume that $\Omega(s)$ is Gaussian. By a suitable linear
transformation we can always re-define $s$ in such a way that this
distribution has zero mean and variance $1/2$, such that
$\Omega(s) \sim \exp(-s^2)$. Furthermore, we assume $f(x) = \chi
x^2$, such that the resulting Hamiltonian is given as
\begin{equation}
  \mathcal{H} = \chi \sum_{\langle ij \rangle} (s_i - s_j)^2
  + k_BT \sum_i s_i^2 .
  \label{eq:GCHamiltonian}
\end{equation}
The parameter $\chi$ takes on a similar meaning as in regular solution
models, and expresses how strong the penalty between different species
is as a function of $s$. The main interest in this particular choice
is its mathematical convenience. The state located at
position $\vec{r}$ is given via Fourier expansion as
\begin{equation}
   s(\vec{r}) = L^{-d/2}\sum_{\vec{k}} a_{\vec{k}} \exp(i \vec{k} \cdot \vec{r}) ,
\end{equation}
where $a_{\vec{k}} = a_{-\Vec{k}}^*$ is the amplitude associated to
mode $\vec{k}$. The quadratic terms of the Hamiltonian decouple in
reciprocal space, leading to a simple form:
\begin{equation}
  \mathcal{H} = \sum_{\vec{k}} \left\vert a_{\vec{k}} \right\vert^2
  \left(k_B T + \chi b_{\vec{k}}\right) ,
\label{eq:HamiltonianK}
\end{equation}
where
\begin{equation}
  b_{\vec{k}}= \sum_{\vec \delta} \left[ 1 - \cos
    \left( \vec{k} \cdot \vec{\delta} \right) \right] ;
\end{equation}
here $\vec{\delta}$ are the $2 d$ vectors that connect a given site to
its nearest neighbors. Since the sum runs over independent modes, they
are independent.

%\textbf{checked up to here}.

A quadratic dependence directly yields a Gaussian distribution for all
$a_{\vec{k}}$, with variance given by
$\sigma_{\vec{k}}^2 = k_B T (k_B T + \chi b_{\vec{k}})^{-1}$.

The correlation function in reciprocal space directly yields the
Ornstein-Zernike form $G_k \sim (1+\xi^2 k^2)^{-1}$, from which the
correlation length $\xi= (8 \pi \chi / kT)^{1/2}$ can be extracted
\cite{herrmann_statistical_2023}. For low temperatures,
$T \lesssim 8 \pi \chi / L^2$, the system exhibits system-spanning
correlations, leading to a plateau region (see
\cite{girard_finite-size_2021}). Behavior of the system within this
correlated regime is precisely the goal of the present article. To do
so, it is useful to investigate the volume-average variance
$\sigma_L^2$. For a single configuration of the spins, this central
moment can be computed as:

\begin{equation}
  \sigma_L^2 = L^{-d} \sum_i \left(s_i - L^{-d}\sum_j s_j\right)^2
  = L^{-d}\sum_{\vec{k}\neq 0} |a_{\vec{k}}|^2,
\end{equation}

As discussed above, although it is an intensive quantity, for finite
systems, a probability distribution is associated to $\sigma^2_L$. A
closed form expression for the distribution of $\sigma_L^2$ cannot be
found. However, a reliable approximation at low temperatures can be
computed (details in SI). This is derived by considering that mode
amplitude follow Gaussian distributions, and therefore
$a^2_{\vec{k}}$ is distributed as $\mathrm{Gamma}(\frac{1}{2},
2\sigma^2_{\vec{k}})$. The characteristic function can therefore be
employed to express the moment generating function of $\sigma_L^2$.

We now consider a thought experiment, in which we assign living cells 
undergoing homeostasis to the grand-canonical ensemble, resting at $T_0$. 
The volume-averaged composition variance $\sigma_L^2$ is given by a realization
of $\sigma_L^2(T_0)$. We then quench this system to temperature $T$, and wait
until diffusion takes place. However, we assume that chemical regulation takes
place on a much slower timescale than our experiment. Consequently, on this timescale
the system lives in a canonical ensemble, and the volume-averaged composition
variance remains fixed. 

In order to translate this experiment in the language of statistical mechanics,
we create a new ensemble living at temperature $T$, which has the same functional form for internal energy,
but where the volume-averaged variance is constrained. To do so, we introduce a Lagrange
multiplier, heron labeled $g_2$. Since we are constraining the second moment, it
is natural to apply the multiplier to the sum of squares of $s$,
leading to a constrained Hamiltonian:

\begin{equation}
  \mathcal{H}_C = \chi \sum_{\langle i j \rangle} (s_i - s_j)^2
  + g_2(T, T_0, L) \sum_i s_i^2
\label{eq:HamiltonianG2}
\end{equation}

We label the volume-averaged variance in this new ensemble by $\sigma_C^2$. 
We discard for a brief moment any fluctuations of either
$\sigma_L^2$ or $\sigma_C^2$. Given our previous thought experiment, we
therefore have $\langle\sigma_C^2(T) \rangle= \langle\sigma_L^2(T_0) \rangle$. We are
now interested in the behavior of the associated multiplier
$g_2(T, T_0, L)$ away from the trivial solution $g_2(T,T,L) = kT$. In
general, closed-form expressions are only available for $d=2$; however
approximations can be derived in the vicinity of $g_2 = 0$ (see SI).

This system exhibits behavior similar to phase transitions. First, we
make the following observation: for any finite $L$ and $T_0 > 0$,
there exists a value of $T$ such that $g_2(T = T_a, T_0, L) = 0$ (see
SI). We can further interpret this by mapping \eqref{eq:HamiltonianG2}
to its equivalent field representation:
$F[\phi] = \chi |\nabla \phi|^2 + g_2(T, T_0, L) \phi^2$; this mapping
associates an apparent phase transition temperature $T_a$ when
$g_2(T=T_a, T_0, L) = 0$, which can be inverted to obtain
$T_a(T_0, L)$. The reader should be extremely careful of this
interpretation, and that we have added an "apparent" label to this
transition temperature.

Limiting behavior of $T_a(T_0, L)$ is of particular interest. Namely,
$T_a(T_0, L\rightarrow\infty) = 0$ for $d\leq 2$, indicating that this
behavior is not necessarily a thermodynamic transition. Rather, there
is a "proper" transition, with lower critical dimension of 2. More
interestingly, we also have
$\lim_{T_0 \rightarrow 0} T_a(T_0, L) / T_0 \rightarrow 1$,
independently of $d$. We had numerically shown this behavior in
\cite{girard_finite-size_2021}. In particular, we had putatively
linked the existance of this transition to correlation length in the
grand canonical ensemble, $\xi$, which would be commensurate with
system size, such that $\xi \sim L$. For $d < 2$, the leading order
correction to $T_a / T_0$ is $\mathcal{O}(L^2 / \xi^2)$ (see SI),
which is consistent with our previous statements. However, for
$2 < d < 4$, the leading order correction is
$\mathcal{O}(L^{4-d} / \xi^2)$, implying that this behavior can be
achieved with $\xi < L$ in $d=3$.

This thought experiment forms the basis of the phenomenon we want to
properly describe here. Notably, that apparent phase behavior depends
on the statistical ensemble; constraining $\langle \sigma_C^2 \rangle$
as we just did alters the phase behavior. However, $\sigma_L^2$ is
subject to fluctuations, which need to be properly treated.

\subsubsection*{Micro- versus macroscopic thermodynamics}

In the thermodynamic limit, the variance of intensive quantities
decays as $\mathcal{O}(L^{-d})$. This can be straightforwardly derived
by putting into contact multiple systems of size $L \gg \xi$; since
their size is much larger than correlation lengths, one should simply
be able to average intensive quantities. Assuming that their intensive
quantities are well behaved, the central limit theorem guarantees that
the average value is Gaussian distributed, with variance
$\sim L^{-d}$.

The systems under consideration here, have small sizes such that the
total length is comparable to the correlation length, i.e.
$L \sim \xi$. This has the consequence of creating anomalous
fluctuations in the system. Here, we investigate the behavior of
$\sigma_L^2$; this is motivated by two reasons. First, it appears to
directly control the apparent phase behavior, and second it is
mathematically tractable. To compute its properties, we expand the
moment generating function at low temperature. We first note the
unusual form of its mean value:
\[
  \langle \sigma_L^2 \rangle \sim
  \frac{4 - 2^d L^{2-d} }{2d-4} \frac{kT}{\chi}
  + \mathcal{O}\left(\frac{kT}{\chi}\right)^2
\]
This quantity is divergent with increasing $L$ for $d \leq 2$. The
series is obviously only valid for small systems, and the
thermodynamic limit cannot be directly taken from this expansion. This
highlights that what we consider a "small system" is highly dependent
on the details, namely temperature. The fluctuations of $\sigma_L^2$
for $d < 2$ diverge at the same rate with $T$ as its mean value,
hinting that the coefficient of variation, $I_d$, is a good quantity
to investigate. This quantity can be approximately calculated (see SI)
as:
\begin{equation}
  I_d = \frac{\sqrt{\mathrm{Var}(\sigma_L^2)}}
  {\langle \sigma_L^2 \rangle}
  \approx
  \frac{2^{\frac{d}{2}} (d-2) \pi^{-\frac{d}{4}}}
  {4 L^d - 2^d L^2}
  \sqrt{\frac{\Gamma\left(\frac{d}{2}\right)(16 L^d - 2^d L^4)}{d-4}}
\label{eq:Invariant}
\end{equation}
This function has no dependence on $L$ for $d<2$. Power laws are
recovered for $d>2$, e.g. $I_3 \sim L^{-1}$, while a logarithmic
divergence is obtained for $I_2$. Usual thermodynamic scaling $L^{-d}$
is recovered for $d>4$. We postulate that this behavior is connected
to the critical dimensions of the transition. The transition has lower
critical dimension of 2 (see above and SI). The XY model is equivalent
to \eqref{eq:GCHamiltonian} at low temperatures, and the former has
upper critical dimension of 4. Similarly, leading order corrections to
$T_a$ in $L$ vanish for $d>4$.

In our previous thought experiment, $\sigma_L^2$ controlled the
apparent demixing temperature; more precisely, we have
$T_a \propto \sigma_L^2$ (see SI). If the thought experiment is
repeated multiple times, then $I_d$ can be directly related to a
measurable quantity: the coefficient of variation of the apparent
transition temperature over the realizations.

Asymptotic behavior of $I_2$ is not universal across all moments of
$\mathcal{D}$. Specifically, the volume-averaged skew, $\gamma_L$ does
not appear to follow the same functional dependency. While we are
unable to express its moment generating function, a numerical
investigation shows that $\mathrm{Var}(\gamma_L) \sim \log(L)^{-3}$ in
$d=2$ (not shown). This will become important in the later discussion,
where $\sigma^2_L$ will take the role of effective temperature and
$\gamma_L$ of composition (magnetization).

% This will be particularly important for transitions, as $\gamma_L$
% takes the role of magnetization for this system. With this in mind,
% the picture we are making of our thought experiment, is that the
% variance $\sigma_L^2$ effectively controls temperature, while
% $\gamma_L$ controls composition, and these two quantities have
% different fluctuations associated with them.

\subsection*{Gaussian model as a general prototype}

Before tacking further questions of phase transition, it is worthwhile
to consider how general the Gaussian system is, in this finite-size
regime. We consider a system in a grand canonical ensemble, with
hamiltonian given by \eqref{eq:GCHamiltonian}. We consider three
distributions for $\Omega$: Uniform, Gaussian, and Hyperbolic Secant;
this particular choice is simply motivated by numerical convenience,
and to have both a leptokurtic and a platykurtic distribution. For
similar reasons, we also consider three functional forms for $f(x)$:
$x$, $x^2$, and $x^4$. Analytically solving \eqref{eq:GCHamiltonian}
can only be done for the Gaussian system, so we turn here to
simulation by means of Monte-Carlo simulations (see SI for
details). We consider here systems with $d=2$, akin to the simulations
in \cite{girard_finite-size_2021}. In addition to the cofficient of
variation $I_2$, we also measure the ensemble averaged kurtosis,
$\kappa(\mathcal{D}) = \langle s^4 \rangle / \langle s^2
\rangle^2$. In the high temperature limit, interactions are
insignificant, and we expect system behavior to be determined by
$\Omega$, such that $\kappa(\mathcal{D}) =
\kappa(\Omega)$. Specifically for the Gaussian system, the analytical
solution of \eqref{eq:GCHamiltonian} shows that
$\kappa(\mathcal{D}) = 3$ at all temperatures.

\begin{figure}
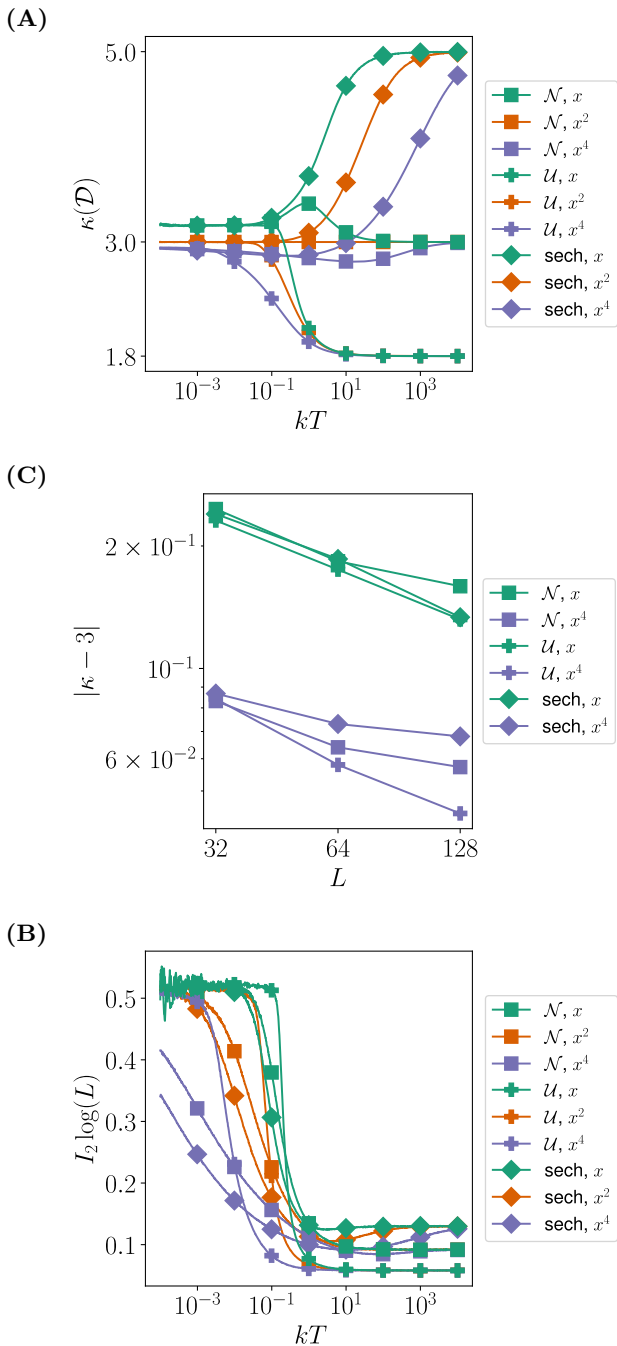

\def\stackalignment{l}
    \centering
    \topinset{\bfseries(A)}{\includesvg[width=0.45\textwidth]
      {Figures/Kurtosis.svg}}{0pt}{-15pt}
    \topinset{\bfseries(C)}{\includesvg[width=0.45\textwidth]
      {Figures/KurtosisVsL}}{0pt}{-15pt}
    \topinset{\bfseries(B)}{\includesvg[width=0.45\textwidth]
      {Figures/Invariant2.svg}}{0pt}{-15pt}
    \caption{Convergence of all systems at low temperatures for $d=2$;
      unless otherwise noted, data is for $L=128$ A) The kurtosis
      varies from the underlying kurtosis of $\Omega$ at high
      temperatures, to a value of $\approx 3$ at low temperatures. B)
      The excess kurtosis $\kappa - 3$ converges towards $0$ as the
      system size increases. C) The quantity $I_2$ approaches a common
      value at low temperatures }
    \label{fig:ConvergenceToGaussian}
\end{figure}

The comparisons of $I_2$ and $\kappa$ for all distributions and
interactions are shown in Fig. \ref{fig:ConvergenceToGaussian}. At low
temperatures, observed values for $\kappa$ are all near $3$. Further
investigating this value shows that the excess kurtosis ($\kappa - 3$)
appears to vanish with increasing $L$. The rate at which systems
converge towards Gaussian is unclear, but could explain the presence
of unknown exponents in \cite{girard_finite-size_2021}. The
coefficient of variation $I_2$ also appears to be unique at low
temperatures. Taken together, this supports the hypothesis that low
temperature behavior is universal. However, what constitutes a low
temperature, and how fast this regime is reached, appears to be highly
dependent on system details.

\subsection*{Apparent universality classes}

Having established that the Gaussian model is a good prototype model
for the finite-size regime, we now turn our attention back to the
nature of these apparent phase transitions. Our constrained
Hamiltonian \eqref{eq:HamiltonianG2} supports an apparent critical
point at $g_2(T_a, T_0) = 0$. We associate to this apparent critical
point, an apparent reduced temperature $\tau_a = (T- T_a) / T_a$. The
system defined by \eqref{eq:HamiltonianG2} has the advantage of being
a Gaussian system, and therefore analytically solvable. In particular,
expanding $g_2$ near $\tau_a = 0$, substituting into the
Ornstein-Zernike form, and extracting the correlation function
directly yields $\xi \sim \tau_a^{-1/2}$ or apparent $\nu =
1/2$. Since the reciprocal space modes are decoupled, the heat
capacity is similarly trivial to compute, and does not show
interesting behavior near $\tau_a = 0$, yielding an exponent
$\alpha = 0$. These values are consistent with the mean-field
universality class. This is not particularly surprising as our field
formulation belongs to the mean-field class. We now want to
investigate whether the concept of universality class translates over
to apparent transitions driven by our finite-size effects.

Extracting these apparent exponents from simulations is not trivial,
and two distinctions from regular thermodynamic transitions complicate
the analysis. First, the apparent critical temperature $T_a$ derives
from $\sigma_L^2$, which shows anomalous fluctuations that are
unbounded for $d \leq 2$. In the formulation of
\eqref{eq:HamiltonianG2}, the value of $g_2$ only prescribes the mean
value $\langle \sigma_C^2 \rangle$. Second, the ensembles given by
\eqref{eq:GCHamiltonian} and \eqref{eq:HamiltonianG2} are not strictly
equivalent, with transitions only appearing in the second. We want to
point out that Fisher renormalization (Fischer 1968, \& Dohm 1974) is
not involved here. Fisher renormalization involves correction in
factors of $1-\alpha$, while $\alpha = 0$ for both
\eqref{eq:GCHamiltonian} and \eqref{eq:HamiltonianG2} near $T=0$ or
$\tau_a = 0$.
 
We investigate a canonical ensemble, where the spin distribution
$\mathcal{D}$ is fully constrained, and where updates are done through
Kawasaki moves. We now need to postulate a distribution $\mathcal{D}$
\textit{a priori}. Here, we use choose deterministic distributions,
generated by $\Phi^{-1}((-1, 1)_L)$, where $\Phi^{-1}$ is an inverse
cumulative distribution, and $(-1,1)_L$ is a linear range of size
$L^d$ which excludes $\pm 1$. This generates a set of numbers with
density commensurate with the cumulative distribution $\Phi$. In the
grand canonical ensemble, our system generates on average a Gaussian
distribution. However, any realization has a non-zero skew, with
magnitude decreasing with $L$. We therefore employ skew-normal
distribution for $\Phi$. This allows us to smoothly change our
distribution from a Gaussian (zero-skew) to one with non-zero
skew;unless otherwise noted skews are chosen to be zero, commensurate
with Gaussian distributions.

To determine universality class, we employ finite-size scaling. We
assume here without justification that the scaling Ansatz applies to
our apparent transitions. In order to proceed, we need to characterize
the system by an order parameter. In order to establish an order
parameter, we turn to distribution of mode power in $k$-space; with
power in a single mode given by $|a_k|^2$. This is motivated by the
observation that total mode power is conserved, and phase separation
is represented by a change in low-$k$ mode power. As our order
parameter needs to match lattice symmetry (translation and $C_4$
rotation), we choose as order parameter squared
$\psi^2 = \langle |a_{x,m}|^2 + |a_{y,m}|^2 \rangle$, where
$a_{x,m} = a_{2\pi \hat{x} / L} $. The susceptibility is given by
$\chi_\psi = \beta \mathrm{Var}(|\psi|)$, and the associated Binder
cumulant by
$U_L = 1 - \frac{1}{3}\langle \psi^4 \rangle / \langle \psi^2
\rangle^2$.

\subsubsection*{Finite-size scaling}

\begin{figure}
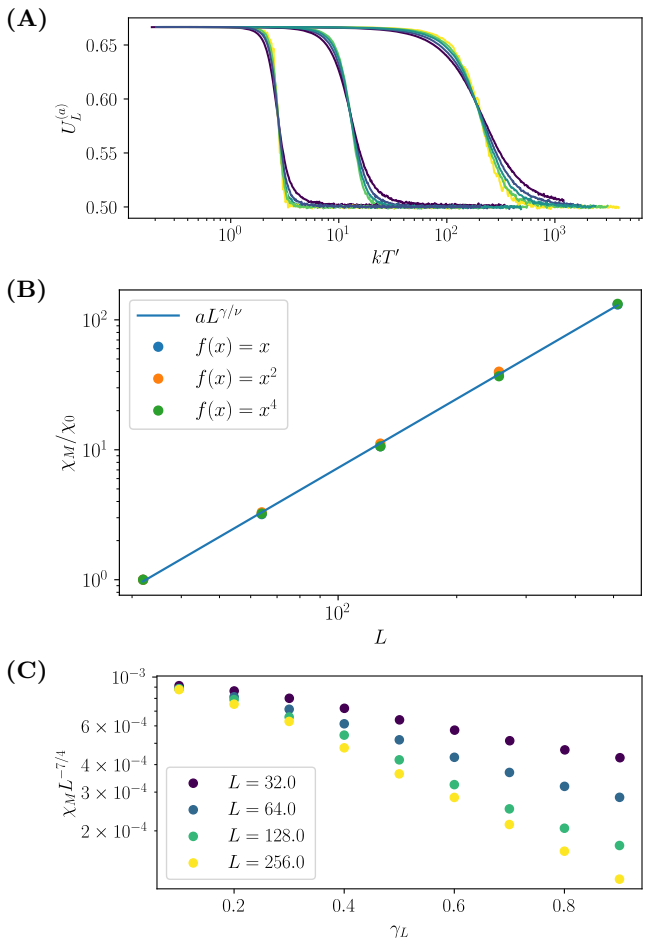

\def\stackalignment{l}
    \centering
    \topinset{\bfseries(A)}{\includesvg[width=0.45\textwidth]
      {Figures/BinderCumulants.svg}}{0pt}{-15pt}\\
    \topinset{\bfseries(B)}{\includesvg[width=0.45\textwidth]
      {Figures/ChiMScaling.svg}}{0pt}{-15pt}\\
    \topinset{\bfseries(C)}{\includesvg[width=0.45\textwidth]
      {Figures/Skew.svg}}{0pt}{-15pt}
    \caption{Finite-size scaling for a $d=2$ system in a canonical
      ensemble for a zero-skew distribution. a) For a given
      interaction, Binder cumulants show a common intersection. from
      left to right: $f(x) = x$, $f(x) = x^2$ and $f(x)=x^4$. b) Peak
      susceptibility shows power-law scaling, independently of $f(x)$,
      with fitted scaling $\chi_M \sim L^{-1.76 \pm 0.01}$. c) Skew
      normal distribution lack common scaling of peak susceptibility}
    \label{fig:2dFSS}
\end{figure}

We only investigate $d=2$ systems, with different forms for the
internal energy, using $f(x) = x^P$, with $P = 1, 2, 4$. The
transition temperature depends on the system size. Upon rescaling
$T' = T \log(L)^{P/2}$, the Binder cumulant curves intersect at a
fixed temperature (see Fig \ref{fig:2dFSS}A). For $P=2$, this
corresponds to the temperature scaling expected based on our Lagrange
multiplier approach. However, we cannot analytically confirm the
scaling for other values of $P$. Numerical investigation of the
susceptibility allows direct extraction of the ratio
$\gamma / \nu = 1.76 \pm 0.01$, consistent with 2D critical Ising
exponents. This value is independent of the choice of $P$.

Based on our experiments for partially constrained ensembles, we
postulate that $\gamma_L$ plays a similar role to magnetization in
usual Ising systems. By using non-zero values of $\gamma_L$ we can
therefore investigate "constant magnetization" ensembles (see Fig
\ref{fig:2dFSS}C). Peak susceptibility decreases with increasing skew,
as one would expect from constant magnetization simulations. Further
investigation of the Binder cumulant and calorimetic curves indicates
that the transition is weakly first order for non-zero skew (not
shown), which is consistent with the mean-field picture of
\cite{graf_thermodynamic_2022} and constant magnetization
simulations. As a simple representation, it is therefore convenient to
associate $\sigma_L^2$ with $T_c$, and $\gamma_L$ with magnetization;
with the latter being associated with \textit{how far} from a critical
point the composition is.

% Since $\gamma_L$ has large fluctuations for $d \leq 2$, on therefore
% does not expect to see hints of critical behavior.

\subsection*{Discussion}

The essential part of this article, is that the notion of ``small''
versus ``large'' systems can be different for multicomponent systems
(continuous spins). While we can give a rigorous definition for
Gaussian systems, it unfortunately depends on details: interactions
and distribution of components. An interesting prediction, is that
maintaining a small system near an apparent phase transition is
relatively simple task: one has to provide a reservoir of many
distinct components, at fixed chemical potential. This could
effectively be implemented by chemical reaction networks, without the
needs of any feedback from the system state. Biological systems
inherently involves such (complex) reaction networks in homeostasis
processes. Consistently, it provides multiple examples of
critical-like behavior, for instance in vesicles extracted from plasma
membranes (GPMV), or within the nuclear pore complex (see
\cite{yu_visualizing_2023,
  honerkamp-smith_introduction_2009,shaw_critical_2021}).

Practically, determining whether an apparent transition in a finite
system is a real (thermodynamic) transition, or a finite-size effect
is a hard problem. The system could also be located near a cross-over
regime. Experiments conducted on the whole system ($L > \xi$) would
not show any transition, while experiments on small scales ($L < \xi$)
would, yielding confusing results. This provides a convenient
\textit{a posteriori} explanation for discrepancies in membranes,
where demixing is observed for GPMVs but not for cellular membranes.

In addition, the interpretation of $s$ in our simple models remains a
problem. We have here ascribed its meaning to represent the chemical
nature of interactions, and assumed that the distribution of $s$ is
fixed in the canonical ensemble. However, interactions in reality are
much more complex, and usually non-additive. A more realistic choice
for $s$ would include both the chemical nature of components, as well
as physical properties, e.g. nematic order parameter. Physical
properties would however not be constrained in a canonical ensemble,
leading to only \textit{some} dimensions of $s$ being constrained. The
precise effect of such constraints remains to be elucidated.

Taking together these results, this simple toy model, which is not
particularly realistic, raises interesting challenges to biology and
physics. Our results imply that the interactions between specific
components would reveal much less information than usually
believed. This kind of measurement gives indications about the value
of $\chi$ in \eqref{eq:GCHamiltonian}, but the overall behavior is
dictated by a competition between entropy ($\Omega$) and enthalpy
($\chi$); How to relate \textit{in-vitro} data to function may be
harder than we currently think.

\subsection*{Author Contributions}

M. Girard designed the research. M. Girard and B. D{\"u}nweg supervised the project. 
M. Girard and F. Herrmann carried out the simulations and analyzed the
data. M. Girard and B. D{\"u}nweg wrote the article.

\subsection*{Acknowledgments}

We acknowledge usage of computational ressources from the Max-Planck
Computing and Data Facilities (MPCDF).

% Uncomment if using bibtex (default)
\bibliography{f_references,f_references_group}

\subsection*{Availability}

Source code and data is available at \url{doi:}

% To ensure reproducibility, source code is available in the SI, as
% well as all files used to produce data.

% Uncomment if using biblatex
% \printbibliography

\section*{Annex}

\subsection*{Simulation methods}

The simulation engine used here is the same as in
\cite{girard_finite-size_2021}. The simulation makes use of a
checkerboard decomposition, and long-range Kawasaki moves for
canonical ensembles. In addition, replica exchange is used. The
calculation is done on graphical processing units.

\subsection*{Canonical sample generation}

To deterministically generate samples from a distribution
$\mathcal{D}$, we generate $\Phi_{\mathcal{D}}^{-1}(\vec{u})$, where
$\vec{u} = (-1 + \delta, -1 + 2\delta + ..., 1-\delta)$,
$\Phi_{\mathcal{D}}(x)$ is the cumulative distribution function, and
$\delta$ is chosen so that $\vec{u}$ contains $L^d$ points. For
$L\rightarrow\infty$, this reproduces all moments of $\mathcal{D}$.

Since we are dealing with finite systems, we strictly enforce that the
sample kurtosis is equal to $3$. To do so, we rescale the generated
distribution by : $f(s) = \mathrm{sgn}(s) |s|^q$, where $q$ is
numerically computed. For the skew normal distribution, we strictly
enforce the sample skew. This is simply done by adjusting the shape
parameter $\alpha$ until the sample skew is attained.

\section*{Supplementary information}

\subsection*{Statistical properties of distributions}

The statistical distributions are chosen for two reasons. First, they
are numerically convenient as it is easy to generate number from such
distributions in a high performance context. Second, it is possible to
order their statistical moments. For equal variance, the n-th moment
of a uniform distribution is smaller than that of a Gaussian
distribution, whereas the hyperbolic secant is higher than the
Gaussian distribution. All odd moments are zero.

\subsubsection*{Uniform}

The uniform distribution between $(-1, 1)$ has a variance of
$1/3$. The associated chemical potential expansion is non-analytical
$\mu(s) \sim \lim_{m\rightarrow\infty} |s|^m$.

\subsubsection*{Gaussian distribution}

The Gaussian distribution is chosen with mean 0 and variance 1. The
associated chemical potential is $\mu(s) = -s^2$.

\subsubsection*{Hyperbolic secant}

The Gaussian distribution is chosen with mean 0 and variance 1. Its
probability density function is:

\[
p(x) = \frac{1}{2}\mathrm{sech} \left(\frac{\pi}{2}x\right)
\]

% Its normalized higher moments are given by:

% \[
%   \bar{M}_{2m} = \frac{1}{4
%   \pi^{2m+1}}\Gamma(2m+1)\left(\zeta\left(2m+1, \frac{1}{4}\right)
%     -\zeta\left(2m+1, \frac{3}{4}\right) \right)
% \]

% Where $\zeta(s, a) = \sum_k (k+a)^{-s}$ is the Hurwitz Zeta
% function. The inverse cumulative function of the distribution can be
% easily expressed as
% $$F^{-1}(x) =
% \frac{2}{\pi}\log\left(\tan\left(\frac{\pi}{2}x\right)\right)$$.

The associated chemical potential is
$$\mu(s) = \sum_m - \frac{(2^{2m} - 1) \pi^{2m} B(2m)}{2m (2m)!}s^{2m}
= - \frac{(\pi s)^2}{8} + \frac{(\pi s)^4}{192} + ...$$, where $B(2m)$
is the $2m$-th Bernoulli number.

\subsection*{Expressions for volume-averaged
  moment $\sigma_L^2$}

For this derivation, we employ real-valued reciprocal representations, i.e. sines and cosines, so that $a_k$ is real. The Hamiltonian \eqref{eq:HamiltonianK} implies that the mode power
$a_k^2$ is distributed as:

\[
p(a_k^2) \sim \mathrm{Gamma}(1/2, 2 \sigma_k^2)
\]

where $\sigma_k^2 = 2 kT (kT +\chi b_k)^{-1}$. For the sake of
clarity, we omit the Boltzmann constant and the variable $\chi$ from
further derivations, implicitly folding it into $T$, which should be
interpreted as $kT / \chi$. Since $\sigma_L^2$ is a sum over
$\vec{k}$, this can be expressed as the product of moment generating
function, computed as
$\mathcal{M}(f) = \mathrm{E}\left[\exp(t f)\right]$. From the Gamma
distribution, we have:

\[
 \mathcal{M}(a_k^2) = \frac{1}{(1-2\sigma_k^2 t)^{1/2}}
\]

And therefore:

\[
\mathcal{M}(\sigma_L^2) = \prod_k \frac{1}{(1-2\sigma_k^2 t / L^{d})^{1/2}}
\]

We first use the identity: $\prod_k f(k) = \exp(\sum_k \log(f(k)))$,
where the sum is then approximated by an integral. We further use the
series expansion $b_k \sim (2\pi)^2 k^2/L^2$ and isotropically
integrate yielding:

\[
  \mathcal{M}(\sigma_L^2) \approx \exp\left(\frac{2
      \pi^\frac{d}{2}}{\Gamma\left(\frac{d}{2}\right)} \int_1^{L/2}
    -\frac{1}{2}\log\left(1-2\frac{\sigma_k^2 t}{L^d}\right) k^{d-1}
    dk\right),
\]

which can be solved in terms of special functions. We define the
function $G_d(z)$ as:

\[
G_d(z) = {}_2F_1 \left(1, 1 + \frac{d}{2}; 2 + \frac{d}{2}; -z\right) 
\]

Where $_2 F_1(a, b; c; z)$ is the usual hypergeometric function
defined as:

\[
_{2}F_1(a, b; c; z) = \sum_{n=0}^\infty \frac{(a)_n (b)_n}{(c)_n} \frac{z^n}{n!}
\]

With the rising factorial $(x)_n = \prod_{k=0}^{n-1} (x+k)$. With this
notation, we have:
\begin{widetext}
\begin{eqnarray*}
  \PA{ \frac{2^d d(d+2)L^2(L^d - 4t)T\Gamma\left(\frac{d}{2}\right)}{\pi^{d/2}}}\log(\mathcal{M}(\sigma_L^2)) = 
  -2L^{2+d}\pi^2 (L^d - 4t) G_d\PA{\frac{\pi^2}{T}}\\
  +2^{3+d}\pi^2 (L^d - 4t) G_d\PA{\frac{4\pi^2}{L^2 T}}
  -2^{3+d}L^d \pi^2 G_d\PA{\frac{4L^{d-2}\pi^2}{T(L^d - 4t)}}\\
  - L^2 \PA{
  2L^{2d} \pi^2 G_d\PA{\frac{L^2 \pi^2}{T(L^d - 4t)}}
  +(d+2)(L^d - 4t) T \PA{
  L^d \log\PA{\frac{(\pi^2 + T)L^d}{L^d (\pi^2 + T) -4 t T}}
  + 2^d \log\PA{1 - \frac{4L^2 t T}{L^d(4\pi^2 + L^2 T)}}
  }
  }
\end{eqnarray*}
\end{widetext}
Isotropic integration yields a geometric prefactor
$2 \pi^{d/2} / \Gamma(d/2)$, which would in reality be determined by
the specific lattice used. Series expansion of $b_k$ is exact in the
limit of $T\rightarrow 0$, as only low-$k$ modes are
populated. Magnitude of the error at higher temperatures depends on
the system dimension, as the density of states at high-$k$ increases
with $d$ as
$k^{d-1}$.
% The integrand has a divergence for low-$k$ modes that dominate the
% result. However for $d>2$, behavior at finite temperature is
% dominated by high-$k$ modes, and we only achieve correct asymptotic
% behavior for $d \leq 2$.

Raw moments of the distribution can be readily computed, with the
$n-$th moment being equal to the $n$-th derivative of $\mathcal{M}$
with respect to $t$ at $t=0$. The reader can consider working with the
logarithm of $\mathcal{M}$:
$\langle \sigma_L^2\rangle = \partial_t \log(\mathcal{M}(\sigma_L^2)
)\left|_{t=0} \right.$. A similar procedure can be employed to compute
moments of $\sigma_C^2$. Expressions for the average values are:

\begin{widetext}
\begin{eqnarray*}
  \frac{T d (d+2) \Gamma(d/2) L^{d+2}2^{d-2}}{\pi^{d/2}}
  \langle \sigma_L^2 \rangle
  =
  (d+2) L^2 (L^d - 2^d) T
  + 2^{d+2} d \pi^2 G_d \PA{\frac{4\pi^2}{L^2 T}}
  - d L^{d+2}\pi^2 G_d\PA{\frac{\pi^2}{T}}
  \\
  \frac{g_2^2 L^{d+2}2^{d-2} d (d+2)
  \Gamma\left(\frac{d}{2}\right)}{\pi^{d/2}}
  \frac{\langle \sigma_C^2 \rangle}{T}
  =  
  (d+2) g_2 L^2 (L^d - 2^d)
  + 2^{d+2} d \pi^2 G_d\PA{\frac{4\pi^2}{g_2 L^2}}
  - d L^{d+2} \pi^2 G_d\PA{\frac{\pi^2}{g_2}}
\end{eqnarray*}
\end{widetext}

\subsection*{The apparent transition temperature $T_a$}

The general expression for $g_2$ can be derived by considering the
above expressions for variances, taking $\langle \sigma_L^2 \rangle$
at temperature $T_0$, setting it equal to $\langle \sigma_C^2 \rangle$
at temperature $T$, and solving for $g_2$. Due to the complexity of
involved functions, we were not able to find closed form
expressions. However, we note that around $g_2 = 0$ (implying
$g_2 L^2 \ll 1)$:

\[
  \frac{\Gamma\PA{1 + \frac{d}{2}} L^d}{d
    \pi^{d/2}}\frac{\langle\sigma_C^2 \rangle}{T} = \frac{4 L^d - 2^d
    L^2}{2^{d+1}(d-2) \pi^2} + \frac{2^d L^4 - 16L^d}{2^{d+3}(d-4)
    \pi^4}g_2 + \mathcal{O}(g_2^2)
\]

The integration to obtain this expression is only valid at low temperatures, the integrand is strictly monotonic with $T$, and therefore $\langle\sigma_C^2 \rangle$ is, in general, a strictly monotonic function of $T$, with the limit $\lim_{T\rightarrow 0} \langle\sigma_C^2 \rangle = 0$. The variance $\langle\sigma_C^2 \rangle$ is therefore finite at $g_2 = 0$. This implies that, without loss of generality, there exists a $T_a < T_0$ such that $g_2(T_a, T_0, L) = 0$. 

\subsubsection*{Thermodynamic limit}

We note that the expression for $\langle \sigma^2_C \rangle$ above is
not valid in the thermodynamic limit. Nevertheless, the scaling:

\[
\frac{  \langle\sigma_C^2 \rangle}{T }  \sim \frac{4 L^d - 2^d L^2}{L^d (d-2)} + \mathcal{O}(g_2)
\]

imposes a necessary condition for the transition to survive the
thermodynamic limit: the constant term needs to be strictly
positive. The transition therefore vanishes for $d\leq 2$. For higher
dimensions, we first take the limit:

\[
\lim_{L\rightarrow\infty}\langle\sigma_C^2\rangle \sim \frac{(d+2) g_2 - d\pi^2 G_d\PA{\frac{\pi^2}{g_2}}}{g_2^2 }
\]

This function diverges near $g_2 = 0$ for $d \leq 2$, but is finite
for $d>2$. A thermodynamic transition therefore exists iff $d > 2$.

\subsubsection*{Low temperature behavior of $T_a$}

To derive behavior of $T_a$ at low temperatures, we consider the
expansion of $\langle \sigma_C^2 \rangle$ above, along with the
expansion of $\langle\sigma_L^2\rangle$ at low temperatures:

\[
  \langle \sigma_L^2 \rangle \sim
  \frac{d(4L^d - 2^d L^2) \pi^{d/2 - 2}}
  {L^d 2^{d+1} (d-2) \Gamma\PA{1 + \frac{d}{2}}} T_0
  + \mathcal{O}(T_0^2)
\]

Since our ensemble, by construction, has
$\langle \sigma_C^2 \rangle = \langle\sigma_L^2\rangle$, we set them
equal, and solve for $g_2$:

\[
g_2 = \frac{4 (d-4) (4L^d - 2^d L^2) \pi^2 (T-T_0)}{(d-2) (16 L^d - 2^d L^4) T}
\]

From which it is clear that $g_2 = 0$ when $T=T_0$. Within this low
temperature regime, these two temperatures are strictly
equal. Expanding to higher order yields the leading order corrections:

% Clearly once the system enters linear regime the two temperatures
% become strictly equal. The leading order correction is:

\[
\frac{T_a}{T_0} = 1 - \frac{(d-2)(16L^d - 2^dL^4)}{4 (d-4) (4L^d -2^d L^2)}T_0 
\]

For a Gaussian system, the temperature $T_0$ is directly linked to the
correlation function, i.e. $T_0 \sim \xi^2$.

\subsection*{Other ensemble constructions}

In general, the nature of apparent critical exponents depends
intimately on the ensemble construction. To demonstrate this, we
constrain a set of volume-averaged moments $c_n = m_n(\mathcal{D})$,
and rely on constrained Monte-Carlo simulations to obtain data. We
represent these constraints by the set $\{c_1, ... \}$. Without loss
of generality, we set $c_1 = 0$, $c_2 = 1$. A canonical ensemble is
reached when the number of constraints is equal to $L^d$. We will
investigate two systems, the iso-variant ensemble $\{c_1 = 0, c_2 = 1\}$
(iso-V), and iso-(variant, kurtosis) $\{c_1 = 0, c_2 = 1, c_4 = x\}$ (iso-K).

It is natural for the partially constrained simulations to use
$\gamma_L$ to characterize this system as it is the lowest
unconstrained moment. The reader will note that odd moments can be
interpreted as deformations of the probability density. At high
temperature, a symmetric distribution will arise, such that any odd
moments are zero. At low temperature, the distribution will become
asymmetric, i.e. skewed, to minimize the energy. Based on our choice,
the susceptibility is then defined as
$\chi_M = L^d \beta\mathrm{Var}(|\gamma_L|)$, and the associated
Binder cumulant is
$U_L = 1 - \frac{1}{3} \langle \gamma_L^4 \rangle / \langle \gamma_L^2
\rangle^2$.

To apply constraints, we select independent quadruplets
$\mathcal{Q} : \{x,y,z,w\}$ using a checkerboard decomposition. We
constrain the sum, sum of squares, and optionally the sum of fourth
powers. This leads to either 1 or 2 degrees of freedom left in the
quadruplet. We then compute gradients of the energy, transform it into
the equivalent of a force, and use force-bias Monte-Carlo to generate
a biased trial move. The choice of "timestep" $A$ is iteratively
selected to maximize the product of $A$ and the number of accepted
moves. The checkerboard decomposition iteratively selects points that
are $2^\ell$ apart with $\ell = 0, 1, ..., \log_2(L) -
2$. Permutations of the quadruplet are also attempted, which results
in long-range diffusion. The different temperatures are coupled
through parallel tempering, with $2^{10}$ different temperatures used
in all simulations, spaced logarithmically.

\begin{figure}
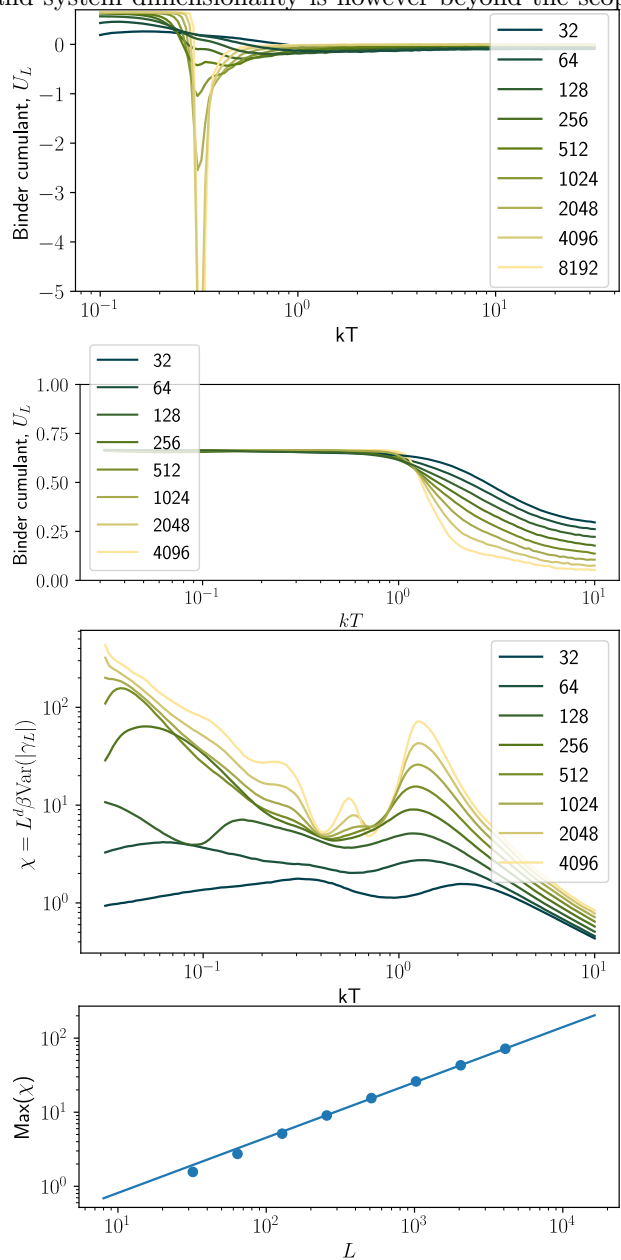

    \centering
    \includesvg[width=0.45\textwidth]{Figures/BinderCumulant_1D_IsoVariant.svg}
    \includesvg[width=0.45\textwidth]{Figures/BinderCumulant_1D_IsoK.svg}
    \includesvg[width=0.45\textwidth]{Figures/Susceptibility_1D_IsoK}
    \includesvg[width=0.45\textwidth]{Figures/ScalingMaxChi_1D_IsoK.svg}
    \caption{Binder cumulant for a $d=1$ system, a) with constraints
      $\{ c_2 = 1\}$. The cumulant does not intersect at a fixed
      tempeture and shows a dip characteristic of first-order
      transitions and b) with constraints $\{c_2 = 1, c_4 =
      3.5\}$. The change in nature of the transition is clear from the
      disapearance of the dip between a) and b). c) Susceptibility of
      the $d=1$ system with $\{ c_2 = 1, c_4 = 3.5\}$. d) The maximum
      value of the susceptibility allows extraction of exponent
      $\gamma / \nu = 0.745 \pm 0.006$ by a linear fit (line).}
    \label{fig:isovariant_binder}
\end{figure}

We first investigate the system $\{ c_2 = 1\}$ in $d=1$ with
$f(x) = x^2$.  The Binder cumulant curves do not intersect at a fixed
temperature, and show a characteristic negative value near the
transition (see Fig \ref{fig:isovariant_binder}). This behavior is
characteristic of first-order transitions, and consistent with the
order parameter probability distribution (not shown). We therefore
assign a first-order type of behavior to this ensemble. It might be
surprising to the reader that a $d=1$ system exhibits a transition,
however, constraining the mean value of a stochastic quantity is
intrinsically different than constraining the quantity itself. The
latter is \textit{effectively} a long-range interaction. Namely,
change in the value of power in one mode, i.e. a single $|a_k|^2$, is
felt by all other modes since their total power is constrained. As
temperature is lowered, the system energy is minimized by condensing
the mode power in a low value of $k$.

The $\{ c_2 = 1, c_4 = x\}$ system with $d=1$ is consistent with a
second-order transition, as can be readily seen on Fig
\ref{fig:isovariant_binder}B. The associated susceptibility increases
with system size (Fig. \ref{fig:isovariant_binder}C, D). Additional
peaks seen in the susceptibility are likely associated with further
transitions similar to \cite{girard_finite-size_2021}, which we will
not characterize here. For all systems, we have that the exponent
$\alpha = 0$ from heat capacity (not shown). However, the value of
other critical exponents continuously varies, ranging from
$\gamma / \nu \approx 0.66$ at $x = 2.5$ to
$\gamma / \nu \approx 0.75$ at $x=3.5$.

We note that both of these ensembles yield a first order transition
with $d=2$. Whether there is a systematic connection between apparent
order, number of constraints and system dimensionality is however
beyond the scope of the current article.

\end{document}